\documentclass[aps,prl,twocolumn,groupedaddress,showpacs]{revtex4}
\usepackage{graphicx}

\begin{document}


\title{Preparing a highly degenerate Fermi gas in an optical lattice}


\author{J. R. Williams, J. H. Huckans, R. W. Stites, E. L. Hazlett, and K. M. O'Hara}
\email[]{kohara@phys.psu.edu}
\affiliation{Department of Physics, Pennsylvania State University,
University Park,\nolinebreak \,Pennsylvania 16802-6300, USA}


\date{\today}

\begin{abstract}
We propose a method to prepare a sample of fermionic atoms in a
three-dimensional (3D) optical lattice at unprecedentedly low
temperatures and uniform filling factors.  The process involves
adiabatic loading of atoms into multiple energy bands of an optical
lattice followed by a filtering stage whereby atoms from all but the
ground band are removed.  Of critical importance is the use of a
non-harmonic trapping potential, taken here to be the radial profile
of a high-order Laguerre-Gaussian laser beam, to provide external
confinement for the atoms.  For realistic experimental parameters,
this procedure should produce samples with temperatures
$\sim10^{-3}$ of the Fermi temperature. This would allow the
investigation of the low-temperature phase diagram of the
Fermi-Hubbard model as well as the initialization of a high-fidelity
quantum register.
\end{abstract}


\maketitle

Investigations of degenerate Fermi gasses loaded into optical
lattices have indicated that these systems are ideal for creating a
robust quantum register for quantum computing applications
~\cite{ZollerFiltering, SmerziQReg} as well as providing a testing
ground for paradigm models of condensed matter physics. Models
currently under investigation include studying Fermi surfaces and
band insulator states ~\cite{EsslingerFermiSurf}, fermionic
superfluidity in a lattice ~\cite{KetterleOLSuperfluid}, and
transport properties of interacting fermions in one and three
dimensional optical lattices ~\cite{InguscioInsulating,
EsslingerTransport}. These seminal experiments demonstrate the high
precision and versatility available in simulating solid state
systems with fermions in optical lattices.

Theoretical studies of such systems have predicted that a number of
exotic phases emerge at low temperatures, including quantum magnetic
ordering and possibly {\it d}-wave
superfluidity~\cite{LukinHighTc,ZollerHubbardToolbox}. However,
temperatures low enough to observe exotic phases such as these are
difficult to achieve when optical lattices are loaded with a
harmonic external confining potential. It has been theoretically
predicted~\cite{KohlThermometry} and experimentally
observed~\cite{EsslingerFermiSurf} that fermions adiabatically
loaded into an optical lattice with harmonic external confinement
experience heating for all but very high initial temperatures and
filling factors (number of atoms per lattice
site)~\cite{BlakieHarmonic}.

Alternative methods to prepare fermionic atoms in optical lattices
at low temperatures and/or high uniform filling factors include:
cooling by adiabatic loading into a three-dimensional (3D)
homogeneous trapping potential with high filling
factor~\cite{BlakieAdiabatic}, defect filtering in a state dependent
optical lattice~\cite{ZollerFiltering}, adiabatic
loading~\cite{SmerziQReg} and filtering~\cite{CiracToolbox} of high
entropy atoms from a 1D lattice with harmonic confinement.

\begin{figure}
 \includegraphics{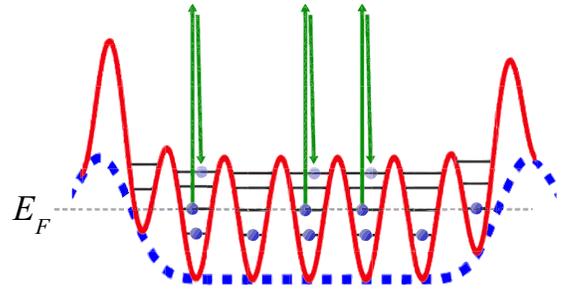}
 \caption{\label{LGFig} (color online) We filter high entropy atoms
 from a combined box-like potential (blue dashed line) and optical lattice
 (red solid line)
 by selectively removing atoms from all but the ground energy band.
 Amplitude modulation of the lattice potential can selectively transfer
 these atoms to high-lying bands via a two-photon transition where they
 then tunnel out of the region.
 Dramatic cooling results when the Fermi energy (prior to filtering)
 lies within the second band.}
 \end{figure}

In this Letter, we propose a method to prepare a highly degenerate
Fermi gas in a 3D optical lattice using a box-like potential for
external confinement and taking advantage of the Pauli exclusion
principle to selectively remove atoms from multiply-occupied lattice
sites. Specifically, we assume that the radial profile of a
blue-detuned, high-order Laguerre-Gaussian (LG) laser beam provides
confinement along each cartesian axis.  The atoms are prepared via a
two step process: (1) adiabatically loading atoms initially confined
in the LG trap into a superimposed optical lattice, followed by (2)
irreversibly filtering atoms from all but the ground energy band
(see Fig.~\ref{LGFig}).  We find that when the Fermi energy of the
system is sufficiently large, such that atoms begin to significantly
populate the second energy band prior to filtering, considerable
cooling is achieved; whereas for lower filling factors heating is
observed.

\begin{figure}
 \includegraphics{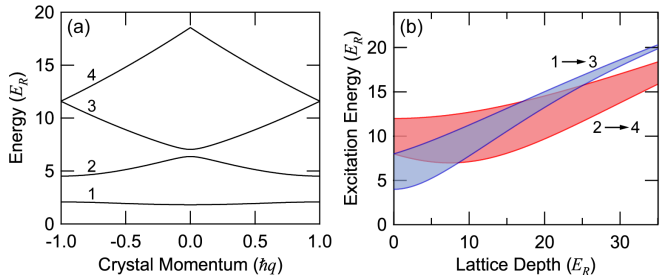}
 \caption{\label{AmplitudeMod} (color online) (a) Band structure for a $5\,E_R$ deep lattice.
 (b) Atoms can be selectively excited from the $n$ to $n+2$ energy band by
 modulating the amplitude of the optical lattice light. The excitation
 energy between bands 1 and 3 (blue) and 2 and 4
 (red) for fermions in all crystal momenta states are shown.
 As the depth of the optical lattice approaches $35\,E_R$,
 the excitation energy bands separate, thereby demonstrating the
 feasibility of performing the band selective excitations required in
 our filtering method.}
\end{figure}

The energy spectrum of a system of ultracold atoms is greatly
affected by the addition of a 3D cubic optical lattice which can be
formed by three perpendicular sets of retroflected Gaussian laser
beams detuned far from resonance. In a homogeneous trapping
potential, the lattice breaks the translational symmetry of the
system, resulting in a series of discrete energy bands whose width
and energy spacings are dependent on the intensity and wavelength of
the laser light (see Fig.~\ref{AmplitudeMod}(a)). Bezett and Blakie
demonstrated that this energy band structure can be exploited to
dramatically increase the degeneracy of the sample for a homogeneous
system~\cite{BlakieAdiabatic}.  For a dense atomic gas, with a
filling factor greater than unity, application of the lattice
increases the Fermi energy, since it lies within the first excited
band, and compresses the Fermi surface resulting in a dramatic
reduction in the degeneracy temperature $T/T_F$, where $T_F$ is the
Fermi temperature.

This band structure also permits state-selective operations to
manipulate and probe the energy distribution of the sample. One such
method involves modulating the depth of the optical lattice to
selectively excite atoms from the $n$ to $n+2$ energy band with no
change in the crystal momentum $q$~\cite{PhillipsLatticeTricks}.  In
contrast to non-interacting Bose systems, where a single $q$ can be
macroscopically occupied, a Fermi system necessarily begins to fill
the bottom band and $\Delta q = 0$ transitions must be excited for
all occupied values of $q$. In Fig.~\ref{AmplitudeMod}(b), we show
band excitation energies as a function of lattice depth for $n=1
\rightarrow 3$ and $n=2 \rightarrow 4$ transitions spanning all $q$
within a Brillouin zone.  By loading the sample into an optical
lattice with a depth of $V_0 = 35\,E_R$ (where $E_R = \hbar^2
k^2/2m$ is the recoil energy and $k$ is the wavenumber of the laser
light), we find that these transitions are well resolved.  It is
therefore possible to apply a filtering process which selectively
removes atoms from all but the ground energy band.   Using adiabatic
rapid passage, population may be selectively transferred from $n=2
\rightarrow 4$ by sweeping the amplitude modulation frequency from
below to above all $2 \rightarrow 4$ transition frequencies while
remaining below the lowest $1 \rightarrow 3$ transition frequency.
Then, lowering the height of the trapping potential allows atoms in
the third and higher energy bands to tunnel out of the system.

In order to experimentally approximate the homogenous lattice
potential described above, we consider the addition of a box-like
external potential produced by a blue-detuned, $\ell^{th}$-order, LG
laser beam with a radial profile
\begin{eqnarray}
V_{{\mathrm{LG}}}(r) & = & V_{\mathrm{peak}}
\left(\frac{2\,e\,r^2}{w_0^2\,\ell}\right)^\ell  e^{-2r^2/w_0^2}
\end{eqnarray}
at the beam waist $w_0$.  For a given charge $\ell$, the peak value
$V_{\mathrm{peak}}$ of the potential occurs at $r_{\mathrm{max}} =
w_0 \sqrt{\ell/2}$ and the width of this peak decreases with
decreasing $w_0$. Therefore, for a given trap size
$r_{\mathrm{max}}$, the LG profile more closely approximates a box
potential when $w_0$ is reduced and $\ell$ is correspondingly
increased.  Trapping of ultracold gases has been demonstrated in
single or crossed beam configurations of LG beams up to $\ell = 16$
~\cite{FatemiLGtrap,DholakiaLGtrap,BashkanskyLGtrap}.

Along a given cartesian axis, we take the single particle
Hamiltonian to be
\begin{eqnarray}
\label{1DHamil}
H(x) &=& \frac{-\hbar^2}{2m}\frac{\partial^2}{\partial x^2} + V_{\mathrm{LG}}(x) \\
& & \quad + V_0 \cos^2(k x + \phi_x) + \frac{1}{2} m \omega^2 x^2,
\nonumber
\end{eqnarray}
where the third term represents a lattice potential of depth $V_0$
and phase offset $\phi_x$. We also include a harmonic term that
arises if red-detuned Gaussian beams are used to produce the lattice
potential; in this case $\omega \propto \sqrt{V_0}$.  The 1D
eigenvalues and eigenfunctions for a given depth of the optical
lattice are calculated by numerically diagonalizing the Hamiltonian
(Eq.~\ref{1DHamil}) using the method described in~\cite{DVRMethod}.
For sufficiently shallow lattice depths, the low energy eigenstates
are delocalized and closely approximate Bloch states in the first
band of a homogeneous system. However, higher energy states are
either localized at the edges of the trap (i.e. near $x =
r_{\mathrm{max}}$) or delocalized and correspond to Bloch states in
higher bands.  While a band structure picture is not strictly valid
for this inhomogeneous system, we classify the set of eigenfunctions
without nodes to constitute the first band,
$\epsilon_{\mathrm{1band},n}$.

We extend this model to three dimensions by assuming a separable
Hamiltonian $H_{\mathrm{3D}} = H(x) + H(y) + H(z)$. For simplicity,
we assume equal lattice depths in each direction. The 3D spectrum
$(\mathcal{E}_{m})$ for a given depth of the optical lattice is then
generated by calculating all possible combinations of the sum
$\mathcal{E}_{m} = \epsilon_{i} + \epsilon_{j} + \epsilon_{k}$ for
all values of the 1D eigenenergies ($\epsilon_{p}$) in each spatial
direction. The 3D energy spectrum for energy states in the first
band of the optical lattice is calculated in a similar manner
$(\mathcal{E}_{{\mathrm{1band}},m} = \epsilon_{{\mathrm{1band}},i} +
\epsilon_{{\mathrm{1band}},j} + \epsilon_{{\mathrm{1band}},k})$.

In calculating thermodynamic quantities during the proposed cooling
method, we assume constant thermal equilibrium before and after the
selective removal of atoms from high-lying bands.  Equilibrium is
maintained by elastic collisions in a 50/50 mixture of spin-1/2
fermions and changes in the trapping potential are adiabatic with
respect to the rethermalization time scale. However, we also assume
that the interactions are weak enough to not significantly modify
the single-particle energy spectrum, $\mathcal{E}_m$. We therefore
use $\mathcal{E}_m$ when calculating the following quantities:
\begin{eqnarray}
\label{ThermoQuantities}
 \nonumber
  N &=& 2\sum_{m}\frac{1}{1+\exp[(\mathcal{E}_{m}-\mu)/k_B T]},\\
  E &=& 2\sum_{m}\frac{\mathcal{E}_{m}}{1+\exp[(\mathcal{E}_{m}-\mu)/k_B T]},\\
     \nonumber
  \frac{S}{k_B} &=& 2 \sum_{m}\ln[1+\exp[(\mathcal{E}_{m}-\mu)/k_B T]]+ \frac{E}{k_B T} -
  \frac{\mu}{k_B T} N,
\end{eqnarray}
where $T$ is the temperature, $\mu$ is the chemical potential of an
atom in either spin state, $N$ is the total number of atoms, $E$ is
the total energy in the system, and $S$ is the total entropy.  The
degeneracy temperature is given by $T/T_F$ where $k_B T_F =
\mathcal{E}_N$, the energy of the $N^{th}$ eigenstate of the
multi-band, 3D spectrum.  After filtering, the thermodynamic
quantities in Eq.~\ref{ThermoQuantities} are calculated for atoms
only in the first band using $\mathcal{E}_{{\mathrm{1band},m}}$.

Our proposed method for cooling the atoms is comprised of (1) an
adiabatic increase in the lattice depth starting from zero, (2) a
non-adiabatic selective filtering of atoms and (3) an optional
adiabatic change to a final lattice depth.  To calculate changes in
thermodynamic quantities during these stages we use the following
procedures. For adiabatic changes of the potential we (1) calculate
$S$ for a given $N$ and initial temperature $T_{\mathrm{i}}$ using
the energy spectrum for the initial potential and (2) numerically
solve for $\mu_{\mathrm{f}}$ and the final temperature
$T_{\mathrm{f}}$ in Eq.~\ref{ThermoQuantities} using the spectrum
for the final potential, assuming $N$ and $S$ are conserved.  In
contrast, for the selective filtering stage we (1) start from a
thermalized sample of $N_{\mathrm{i}}$ atoms at temperature
$T_{\mathrm{i}}$ for a given spectrum, (2) calculate, given this
distribution, the energy $E_{\mathrm{f}}$ and number
$N_{\mathrm{f}}$ for atoms {\emph{restricted to the first band}},
and (3) solve for $\mu_{\mathrm{f}}$ and the temperature
$T_{\mathrm{f}}$ in Eq.~\ref{ThermoQuantities} using the multi-band
energy spectrum assuming the sample equilibrates with total energy
$E_{\mathrm{f}}$ and number $N_{\mathrm{f}}$.

\begin{figure}
 \includegraphics{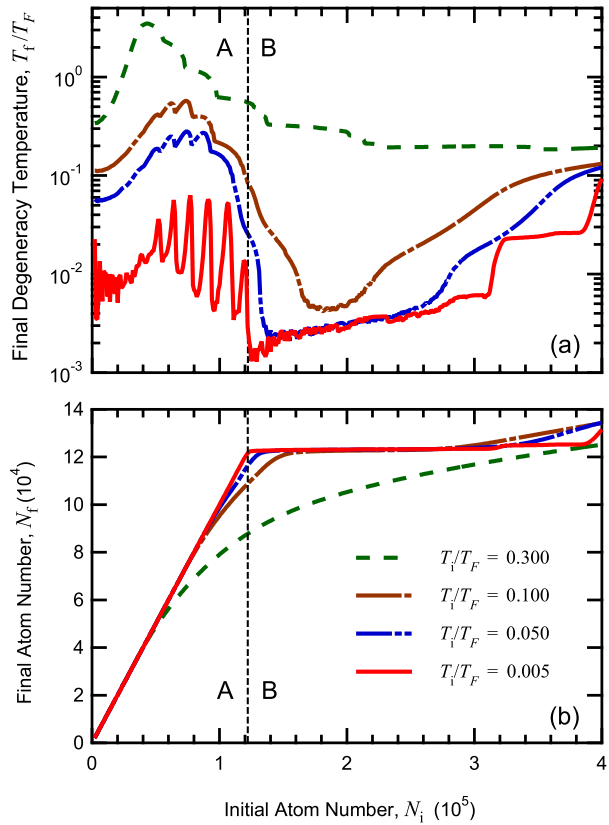}
 \caption{\label{LGTempData} (color online) As a function of
 initial atom number we report (a) the final degeneracy
 temperature and (b) the final atom number after implementing
 the proposed cooling and filtering procedure for various
 initial temperatures between $0.005$ and $0.3\,T_F$.  The vertical dashed
 line represents the number of atoms for which the Fermi energy
 enters the second band.  The trap and lattice parameters are
 as described in the text.}
\end{figure}

We consider a 50/50 spin mixture of $^6$Li atoms initially trapped
in a LG trapping potential with $\ell = 12$, $V_{\mathrm{peak}} =
35\, E_R$ and $r_{\mathrm{max}} = 13.5\,\mu\mathrm{m}$.  For
reasonable lattice beam properties ($k = 2\pi/1064\,\mathrm{nm}$ and
a waist of $200\,\mu\mathrm{m}$) we find $\omega =2 \pi\,(586
\mathrm{Hz})$ for the final lattice depth $V_{0,\mathrm{f}} = 35\,
E_R$.  The final degeneracy temperatures after adiabatic loading and
filtering, along with the final atom number are shown in
Fig.~\ref{LGTempData} for various initial degeneracy temperatures
and sample sizes. In each case $\phi_x = \phi_y = \phi_z = 0$.  This
data shows that the thermodynamic properties of the system are
highly dependent on the initial filling factor and can be separated
into two distinct regions A and B.  The vertical dashed line which
separates the regions represents the number of atoms at which the
Fermi energy enters the second band.

In region A, the Fermi energy before filtering lies below the second
energy band. For very low filling factors, an increase in $T/T_F$ is
observed for all initial temperatures. This increase in $T/T_F$
occurs because $T_F$ decreases more than $T$ as the lattice depth
increases. Additionally, Fig.~\ref{LGTempData}(b) demonstrates that
we are not significantly filtering atoms for low initial
temperatures. As the Fermi energy approaches the second energy band,
we see a dramatic decrease in the final $T/T_F$. In this regime,
where atoms are beginning to significantly occupy the localized
energy states at the edge of the trapping potential, a dramatic
increase of the Fermi energy is observed.

In region B, the density is such that the Fermi energy before
filtering lies within the second band.  In this region the adiabatic
increase of the lattice depth results in a dramatic increase in
$T_F$, a substantial reduction in the temperature
$T$~\cite{BlakieAdiabatic}, and allows for a significant reduction
in entropy during the filtering stage. We find that significant
cooling is achieved for initial temperatures in the vicinity of
$T_{\mathrm{i}} = 0.1\,T_F$. Above this initial temperature cooling
is less efficient. Below this initial temperature, the final $T/T_F$
after filtering saturates. As can be seen in
Fig.~\ref{LGTempData}(b), $N_{\mathrm{f}}$ is extremely insensitive
to fluctuations in $N_{\mathrm{i}}$ for low initial temperatures.
For example, at $T_{\mathrm{i}} = 0.05\,T_F$, a variation of $\pm
10\%$ around $N_{\mathrm{i}} = 1.6 \times 10^5$ yields a variation
of only $+0.09\%/-0.2\%$ in $N_{\mathrm{f}}$.

The cooling efficiency and number filtering were dependent on the
choice of phases $\phi_x$, $\phi_y$ and $\phi_z$ due to the
sensitive effect these phases had on the the location of localized
edge state eigenenergies relative to the Fermi energy. To study this
effect, we modeled the system allowing the phase in each direction
to be independently selected from the set $\phi_\alpha =
(0,\pi/10,...,\pi/2)$.  We considered samples with an initial
temperature $T_{\mathrm{i}} = 0.05\,T_F$ and an initial number
$N_{\mathrm{i}} = 1.6 \times 10^{5}$ atoms, parameters within the
saturated regime for all choices of phase and close to optimal for
cooling. From the set of all possible phase combinations, we find an
average final temperature $T_{\mathrm{f}} = 0.0031\,T_F$ where
$10\%$ of the ensemble achieved temperatures below $T_{10} =
0.0023\,T_F$ and $90\%$ were below $T_{90} = 0.004\,T_F$. From this
same set, we find an average final number $N_{\mathrm{f}} = 1.20
\times 10^{5}$, with $N_{10} = 1.18 \times 10^{5}$, and $N_{90} =
1.22 \times 10^{5}$. The filtering process further results in a
substantial reduction in entropy. The initial entropy per atom
$s_{\mathrm{i}} = 0.28\,k_B$ is reduced to an average final value of
$s_{\mathrm{f}} = 0.024\,k_B$, with $s_{10} = 0.014\,k_B$, and
$s_{90} = 0.033\,k_B$.

It is in general possible to prepare atoms at a low $T/T_F$ in a
{\emph{shallow}} lattice potential, if so desired, by adiabatically
reducing the lattice depth after the filtering stage.  Continuing
the example from above, when the lattice depth is reduced to
$5\,E_R$ we find an average final temperature $T_{\mathrm{f}} =
0.002\,T_F$, $T_{10} = 0.0013\,T_F$, and $T_{90} = 0.0028\,T_F$.

\begin{figure}
 \includegraphics{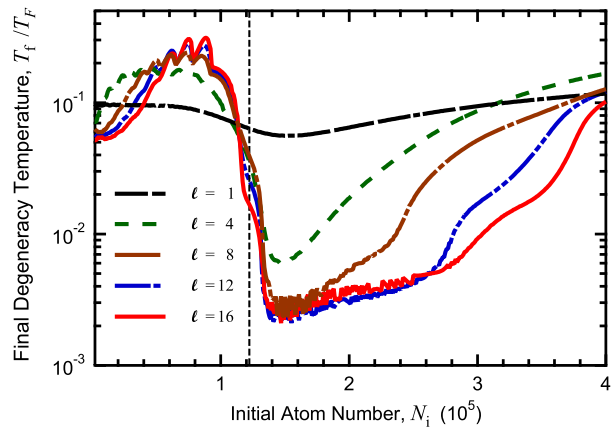}
 \caption{\label{LGLValData} (color online) The effects of the
 charge $\ell$ of the Laguerre - Gaussian trapping potential
 on the efficiency of our proposed cooling and filtering
 method.  For each data set, the initial temperature
 $T_{\mathrm{i}} = 0.05\,T_F$ and the phases
 $\phi_x = \phi_y = \phi_z = 0$. For all $\ell$ values, the number
 of atoms at which the Fermi energy enters the second band
 (vertical dashed line) is held constant.}
\end{figure}

We now consider the effects of the charge $\ell$ of the LG beam for
samples with an initial $T_{\mathrm{i}} = 0.05\,T_F$, phase $\phi_x
= \phi_y = \phi_z = 0$, final lattice depth of $35\,E_R$, and
various initial atom numbers. For each $\ell$-value, the waist of
the LG beam is adjusted such that the number of states below the
second energy band is held constant at $1.22 \times 10^5$.  As shown
in Fig.~\ref{LGLValData}, the cooling efficiency of this procedure
is highly dependent on the charge.  Note that for $\ell = 1$, which
approximates harmonic external confinement, the final degeneracy
temperature $T_{\mathrm{f}}/T_F$ never drops below its initial value
of $T_{\mathrm{i}}/T_F = 0.05$. For $\ell \gtrsim 8$, the minimum
degeneracy temperature saturates to $T_{\mathrm{f}}/T_F \lesssim
0.003$. For higher values of $\ell$, the extent of the saturation
regime grows. We believe that this saturation is caused by localized
atoms at the edges of the LG potential rethermalizing into higher
energy bands.

In this Letter we proposed a method for preparing a sample of highly
degenerate fermions by adiabatic loading into a combined optical
lattice and ``box-like'' trapping potential followed by selective
removal of atoms from all but the ground energy band. Numerical
calculations for sample sizes $\sim 10^5$ predict that temperatures
$\sim 10^{-3}\,T_F$ can be prepared in this manner. This method is
robust against initial number and temperature fluctuations for a
sufficiently cold initial sample of atoms and yields samples with
little variance in the final number. While the selective removal of
atoms must occur in a deep lattice (in order to spectrally resolve
the band excitations), subsequent reduction of the lattice depth, if
desired, yields a modest amount of additional cooling. We expect
that this method can be scaled to larger samples for which still
lower degeneracy temperatures would be attained due to the
diminished role localized edge states would play. Further, the
``box-like'' trapping potential offers an ideal spatial profile for
simulating solid state physics with degenerate atoms in optical
lattices as the relatively flat central region allows for a large
number of delocalized states while the curvature at the edges of the
traps removes the constraint of loading exact atom numbers to
realize insulating states.  Atoms prepared in this manner should be
sufficiently cold to explore quantum spin phases of fermionic atoms
which are currently inaccessible, and could provide a physical
realization of an essentially perfect quantum register.

\begin{acknowledgments}
We gratefully acknowledge support from the Air Force Office of
Scientific Research (Award No. FA9550-05-1-035112) and the Physics
Division of the Army Research Office (Grant No. W911NF-06-1-0398) as
well as discussions with David Weiss regarding this work.
\end{acknowledgments}



\end{document}